\documentclass[submission, Phys]{SciPost}
\usepackage{graphicx} % Include figure files
\usepackage{bm}
\usepackage{txfonts}
\usepackage{mathrsfs}
\usepackage{amsmath}
\usepackage{xcolor}

\usepackage{subcaption}

\def\b0{{\mathbf 0}}

\def\b0{{\mathbf 0}}

\def\eps{\epsilon}

\newcommand{\vkd}{\textbf{k}_d}

\newcommand{\di}{\mathrm{d}}

\newcommand{\epsi}{\frac{1}{\psi}}

\def\beq{\begin{equation}}
\def\eeq{\end{equation}}

\begin{document}

% TODO: write your article's title here.
% The article title is centered, Large boldface, and should fit in two lines
\begin{center}{\Large \textbf{
Thermodynamic Casimir forces in strongly anisotropic systems within the $N\to \infty$ class
}}\end{center}

% TODO: write the author list here. Use initials + surname format.
% Separate subsequent authors by a comma, omit comma at the end of the list.
% Mark the corresponding author with a superscript *.
\begin{center}
M. Łebek\textsuperscript{1*},
P. Jakubczyk\textsuperscript{1},
\end{center}

% TODO: write all affiliations here.
% Format: institute, city, country
\begin{center}
{\bf 1} Institute of Theoretical Physics, Faculty of Physics, University of Warsaw, Pasteura 5, 02-093 Warsaw, Poland
\\
% TODO: provide email address of corresponding author
*m.lebek@student.uw.edu.pl
\end{center}

\begin{center}
\today
\end{center}

% For convenience during refereeing: line numbers
%\linenumbers

\section*{Abstract}
{\bf
% TODO: write your abstract here.
We analyze the thermodynamic Casimir effect in strongly anisotropic systems from the vectorial $N\to\infty$ class in a slab geometry. Employing the imperfect (mean-field) Bose gas as a representative example, we demonstrate the key role of spatial dimensionality $d$ in determining the character of the effective fluctuation-mediated interaction between the confining walls. For a particular, physically conceivable choice of anisotropic dispersion relation and periodic boundary conditions, we show that the Casimir force at criticality as well as within the low-temperature phase is repulsive for dimensionality $d\in (\frac{5}{2},4)\cup (6,8)\cup (10,12)\cup\dots$ and attractive for $d\in (4,6)\cup (8,10)\cup \dots$. We argue, that for $d\in\{4,6,8\dots\}$ the Casimir interaction entirely vanishes in the scaling limit. We discuss implications of our results for systems characterized by $1/N>0$ and possible realizations in the contexts of optical lattice systems and quantum phase transitions.
}

% TODO: include a table of contents (optional)
% Guideline: if your paper is longer that 6 pages, include a TOC
% To remove the TOC, simply cut the following block
\vspace{10pt}
\noindent\rule{\textwidth}{1pt}
\tableofcontents\thispagestyle{fancy}
\noindent\rule{\textwidth}{1pt}
\vspace{10pt}

\section{Introduction}
\label{sec:intro}
% TODO: write your article here.
The thermodynamic Casimir effect received substantial interest over the last years \cite{Mostepanenko_1988, Krech_book, Kardar_1999, Brankov_book, Dantchev_2003, Gambassi_2009, Klimchitskaya_2009, Maciolek_2018} both from theoretical and experimental points of view. The occurrence of these fluctuation-mediated interactions becomes recognized in an increasing number of systems of surprising diversity (such as, for example, biological membranes \cite{Machta_2011, Machta_2012}) and their existence and properties are nowadays firmly established experimentally \cite{Maciolek_2018} on both qualitative and quantitative levels. The validity of the theoretical predictions has also been tested in extensive and impressive numerical simulations (see e.g. \cite{Vasilyev_2009}) .
% (see for example \cite{Vasilyev_2009, Hasenbush_2010 }).

An obviously important basic property of the Casimir force is its sign. According to exact theorems\cite{Li_1997, Kenneth_2006} formulated in the context of the electrodynamic Casimir effect, the fluctuation-induced Casimir force acting between bodies related by a reflection must be attractive.  The same is expected to hold true for the thermodynamic Casimir effect implying attractive character of the thermodynamic Casimir interactions in systems involving identical boundaries (representing identical molecules) immersed in a uniform fluid. 
%and therefore relevant for a broad class of boundary conditions, such as periodic, Dirichlet, von Neumann or Robin. 
The above-mentioned expectation has been confirmed in numerous theoretical studies (both exact and approximate) as well as in simulations. We note in passing that a repulsive Casimir effect was also considered in complementary situations, where the boundary conditions are different on each of the bodies and can be experimentally tuned \cite{Soyka_2008, Nellen_2009}. 

The exact statements of Refs.~\cite{Li_1997, Kenneth_2006} rely however on the explicit form of the field propagator and its quadratic dependence on momentum. In the present paper we explore situations where this condition is not fulfilled. Our analysis indicates that it leads to a far-going deviation from the usual situation, and, in particular, yields the Casimir interaction attractive, repulsive, or zero depending on the system dimensionality. This is completely opposite to the usual cases extensively studied before, where dimensionality has no impact on the force sign. 

The present analysis is carried out implementing a particular microscopic model, the so-called mean-field or imperfect Bose gas on an anisotropic lattice, but the conclusions are relevant to the entire universality class, which may encompass a broad diversity of physical systems.   Our primary motivation for studying Bose systems with dispersions deviating from the quadratic form stems from the recognized tunability of the dispersion relation in anisotropic optical-lattice systems \cite{Greif_2013, Imriska_2014} by the Feshbach resonances. As was discussed in Ref.~\cite{Jakubczyk_2018}, considering a tight-binding type model with at least nearest- and next-to-nearest- neighbour hoppings, one may tune the microscopic parameters so that the quadratic component of the dispersion is cancelled. In anisotropic lattices this can be done independently in each of the $d$ spatial directions, leading to a dispersion which is quartic in $m$ ($m\leq d$) spatial directions and quadratic in the remaining $d-m$ directions. This yields a rich phenomenology involving effective dimensional crossovers in the bulk \cite{Lebek_2020}, but also drastically affecting the interfacial properties. Concerning the Casimir effect this manifests itself in two striking effects: change of the power law governing the decay of the Casimir force as function of the distance (which is accompanied by appearance of a non-universal scale governing its amplitude) and a change of the Casimir force sign. 

At the heart of the theory underlying the phenomenology of Casimir interactions lies the concept of the dimensionless scaling function $\Delta(x)$, describing the variation of the excess free energy density $\omega_s$ upon changing the scaling variable $x\sim D/\xi$, where, in the presently considered setup of a slab (hypercubic) geometry, $D$ is the system extension in one of the directions, while $\xi$ denotes the bulk correlation length.  The (linear) system size $L$ in the remaining directions is assumed infinite ($L/D\to\infty$). The excess free energy density $\omega_s$ is generically related to $\Delta(x)$ via 
\begin{equation}
\omega_s = k_B T\frac{\Delta(x)}{D^{d-1}}
\label{Eq1}
\end{equation}
in the so-called scaling limit, where both $D$ and $\xi$ are large as compared to microscopic scales.
The scaling function  $\Delta(x)$ is universal in the sense that it depends on the bulk universality class and the boundary conditions imposed on the fluctuating medium by the confining walls, but not fine microscopic details of the system. The Casimir force (per unit area) is given by $F=-\frac{\partial \omega_s}{\partial D}$. The scaling function $\Delta(x)$ was computed for a broad variety of systems within exact and approximate analytical approaches as well as numerical simulations \cite{Dantchev_2006, Nowakowski_2008, Nowakowski_2009, Dantchev_2011, Diehl_2012, Hasenbusch_2012, Hasenbusch_2013, Vasilyev_2013, Diehl_2014, Dantchev_2014,  Hasenbusch_2015}. It was also measured experimentally (see Ref.~\cite{Maciolek_2018} for a recent review).

There are few known cases, where Eq.~(\ref{Eq1}) does not apply.  One such situation arises in systems exhibiting strongly anisotropic scale invariance\cite{Diehl_2002}, where the singularity of the correlation function at the phase transition is related to (at least) two correlation lengths $\xi_{\parallel}$ and $\xi_{\perp}$ diverging with different critical exponents so that $\xi_{\perp}\sim \xi_{\parallel}^{\theta_A}$, and the anisotropy exponent $\theta_A\neq 1$. As was demonstrated in Ref.~\cite{Burgsmuller_2010}, the Casimir energy decay exponent $\zeta_0=d-1$ in Eq.~(\ref{Eq1}) becomes in such a situation modified. This is interesting, because, for dimensional reasons, a quantity of dimension [length] must then appear in the corresponding expression for $\omega_s$. This in turn may originate only from the microscopic quantities, thus restricting the universal character of the Casimir interaction. Specifically, for the so-called $m$-axial Lifshitz point,\cite{Hornreich_1975, Chaikin_book} Ref.~\cite{Burgsmuller_2010} predicts that Eq.~(\ref{Eq1}) becomes replaced by 
\begin{equation}
\omega_s = k_B T \frac{\Gamma\Delta_m^d(x)}{D^{\zeta_m}}\;,
\label{Eq2}
\end{equation}
where 
\begin{equation}
\zeta_m = \frac{d-m}{\theta_A}+m-1\;,
\label{Eq3}
\end{equation}
and $\Gamma$ is a \emph{dimensionful} scale factor, deriving from microscopic length scales and therefore non-universal. Equations (\ref{Eq2}) and (\ref{Eq3}) apply to the setup, where the confining walls are oriented perpendicular to one of the $m$ ($m\leq d$) directions, where the inverse propagator deviates from the standard quadratic form and is (up to anomalous dimensions) quartic in momentum.  

In this paper we argue that  the scaling function $\Delta_m^d(x)$ occurring in Eq.~(\ref{Eq2}) is strictly zero for the $N\to\infty$ universality class with $m=1$ and periodic boundary conditions in even dimensionalities $d=2n$, $n\in\{2,\,3,\,4,\,\dots\}$. We consider a microscopic model being a representative of this universality class and analyze the properties of the scaling function $\Delta_m^d(x)$ upon varying dimensionality $d$. By an exact analysis we demonstrate in particular that $\Delta_1^d(x)$ changes sign for each $d=2n$ (and is identically equal zero for $d\in\{4,\,6,\,8,\;\dots\}$). In consequence, the corresponding  Casimir interaction  is repulsive for $d\in (\frac{5}{2},4)\cup (6,8)\cup\dots$ and attractive for $d\in (4,6)\cup (8,10)\cup \dots$. This is in stark contrast to the case of isotropic systems with quadratic dispersion, where (for periodic boundary conditions) the Casimir force is always attractive (in any dimensionality and also for the entire family of $O(N)$ universality classes), as guaranteed by the exact statements of Refs. \cite{Li_1997, Kenneth_2006}. We clarify the character of the Casimir interaction in the peculiar case of $d\in\{4,\,6,\,8,\;\dots\}$ by demonstrating that there is no subdominant contribution to the excess free energy, surviving the scaling limit. From continuity in $1/N$ we argue, that (at least for some values of the scaling variable $x$) %the discussed features are not specific to the strict limit of $N\to\infty$ and 
the sign of the Casimir force also changes at particular (presumably non-integer) values of $d$ provided $1/N>0$ is sufficiently small.   

A substantial technical part of our analysis heavily relies on an earlier calculation presented in Ref.~\cite{Lebek_2020}. In that paper we confirmed the predictions summarized in Eq.~(\ref{Eq2}) and Eq.~(\ref{Eq3}) and calculated the scaling function focusing mainly on spatial dimensionalities corresponding to $d=3$. As we demonstrate in the present analysis, varying dimensionality has a drastic and unexpected impact on the emergent physical picture.  In order to avoid repetitions, we will frequently refer to Ref.~\cite{Lebek_2020} throughout the paper.

The present calculation is exact and is carried out for the imperfect Bose gas, which constitutes a particular microscopic representative of the $N\to\infty$ universality class. More precisely, as established in Ref.~\cite{Diehl_2017} for the isotropic case, the imperfect Bose gas is equivalent to the $O(2N)$ model in the limit $N\to\infty$ and the corresponding scaling functions\cite{Dantchev_1996, Dantchev_2004} for Casimir energy differ by a global factor of two. One may check that (at least for periodic and von Neumann boundary conditions) the form of the dispersion has no impact on the study of Ref.~\cite{Diehl_2017} in the aspects exploring connections between the imperfect Bose gas, the interacting $N$-component  Bose gas in the limit $N\to\infty$, and the classical Landau-Ginzburg $\phi^4$-type theory. In consequence an analogous correspondence holds for the anisotropic situations as well. 

The outline of the paper is as follows: In Sec.~II we discuss the model and summarize the relevant elements of its bulk thermodynamics. Sec.~III contains an analysis of the saddle-point equation. Both Sec.~II and Sec.~III strongly rely on Ref.~\cite{Lebek_2020}. However, to drag the correct conclusions it is necessary to keep track of remainder terms (vanishing in the scaling limit) which constitutes the important extension of Ref.~\cite{Lebek_2020}.    The new, physically relevant results are contained in Sec.~IV, where we analyze the excess free energy varying dimensionality. Sec.~V contains a summary and a portion of technical details of the analysis  is postponed to appendices A and B. 
\section{The mean-field Bose gas} 
We consider the mean field (imperfect) Bose gas governed by the Hamiltonian 
\begin{equation}
\hat{H}=\sum_{\bf k} \epsilon_{\bf k}\hat{n}_{\bf k}+\frac{a}{2V}\hat{N}^2 \;.
\end{equation}
In addition to the standard kinetic component the model contains the repulsive mean-field interaction term $\hat{V}_{mf}=\frac{a}{2V}\hat{N}^2$ ($a>0$), which arises from a long-range repulsive part  $v(r)$ of a 2-particle interaction potential in the Kac limit $\lim_{\gamma\to 0}\gamma^dv(\gamma r)$, corresponding to vanishing interaction strength and diverging range.  This limit is very close in spirit to the rigorous treatment of the van der Waals theory of classical fluids \cite{Kac_1963}. Different aspects of this model were studied in recent years \cite{Davies_1972, Buffet_1983, Zagrebnov_2001, Napiorkowski_2011, Napiorkowski_2013, Diehl_2017, Napiorkowski_2017,  Mysliwy_2019, Dantchev_2020} considering both its bulk and finite-size properties. In particular, for the isotropic continuum case it was established \cite{Napiorkowski_2013, Diehl_2017} that the Bose-Einstein condensation in this model  is a representative of bulk $O(N\to\infty)$ universality class. 

 If the model is considered on a lattice, the dispersion $\epsilon_{\bf k}$ may in general be a complicated function of momentum. For example, for a hypercubic lattice it takes the form 
 \begin{equation}
 \epsilon_{\bf k}=\sum_{\bf x} 2t_{\bf x}[1-\cos ({\bf k x})]
 \end{equation}
  where ${\bf x}$ labels the lattice points and $t_{\bf x}$ are the lattice hopping parameters. Generically, when expanded around ${\bf k}=0$, such a dispersion is quadratic. As was shown in Ref \cite{Jakubczyk_2018}, it is however possible to choose the hoppings so that the coefficient of the quadratic contribution cancels and the corresponding asymptotic behavior of $\epsilon_{\bf k}$ is then quartic, or even higher order in momentum. This tuning procedure can be carried our independently in each of the spatial directions. Moreover, as was demonstrated in the analysis of Ref. \cite{Jakubczyk_2018}, only the low momentum asymptotic form of the dispersion is relevant for the critical singularities. Note that non-universal quantities, such as the critical temperature, are certainly affected by this approximation \cite{Jakubczyk_2018}.    We therefore consider 
\begin{equation} 
\label{asssym}
\epsilon_{\bf{k}} \to \tilde{\epsilon}_{\bf{k}}  = \sum_{i=1}^{d-m} t_0 (k_i A)^2 + \sum _{i=d-m+1}^{d} t(k_i A)^4\;, 
\end{equation} 
replacing the dispersion $\epsilon_{\bf{k}}$ with its low-momentum asymptotic form $\tilde{\epsilon}_{\bf{k}}$ and assuming the hoppings had been chosen so that the dispersion is quartic in $m\leq d$ directions (and quadratic in the remaining). We also assume  $t_0>0$, $t>0$ and introduce 
\beq
\tilde{\epsilon}_{k_1}=t_0 (k_1 A)^2 \;\;\;\;\textrm{and} \;\;\;\;\tilde{\epsilon}_{k_d}=t (k_d A)^4
\eeq
for future reference. The quantity $A$ is a microscopic length, which may be identified with a lattice constant. The bosonic particles are assumed spinless for simplicity. The system is $d$-dimensional and is enclosed in a hypercubic volume $V=L^{d-1}D$, where $L\gg D\gg l_{mic}$ and $l_{mic}$ denotes all the microscopic length scales present in the system. The quantity $D$ measures the system extension in the $d$-th direction along which the dispersion is quartic [see Eq.~(\ref{asssym})]. We impose periodic boundary conditions in all the directions (including the $d$-th one). This choice, often preferable in numerical simulations, is clearly not the most physical one and in our study is dictated mostly by convenience. We  checked that implementing the Neumann boundary conditions modifies some numerical factors, but does not change our major conclusions. We also point out that the presumably most realistic Dirichlet and Robin boundary conditions add  technical complexity to the present study and will not be considered here.   

Below we sketch the essential steps leading to the solution of the model \cite{Napiorkowski_2011, Napiorkowski_2013, Jakubczyk_2018}.   We work within the framework of the grand canonical ensemble. The corresponding grand canonical partition function may be written as\cite{Napiorkowski_2011} 
\begin{equation}
\label{partition}
\Xi(L,D,\mu,T)=-i\exp\bigg(\frac{\beta V}{2a} \mu^2\bigg)\sqrt{\frac{V}{2 \pi \beta a}}\int_{\beta \alpha - i \infty}^{\beta \alpha + i \infty} \di s \, \exp[-V \varphi(s)]\;.
\end{equation}
The parameter $\alpha<0$ is arbitrary, $\beta^{-1}=k_B T$ and 
\begin{equation}
\varphi(s)= \frac{1}{\beta a}\bigg(-\frac{s^2}{2}+s\beta \mu\bigg)-\frac{1}{V}\log \Xi_0\bigg(\frac{s}{\beta},T\bigg)\;
\end{equation}
with the quantity $\Xi_0\big(\frac{s}{\beta},T\big)$ denoting the grand canonical partition function of the noninteracting Bose gas\cite{Ziff_1977} evaluated at chemical potential $\mu=\frac{s}{\beta}$ and temperature $T$. The presence of the volume factor in the term $\exp[-V \varphi(s)]$ in Eq.~(\ref{partition}) guarantees that the saddle point analysis of Eq.~(\ref{partition}) becomes exact for $V\to\infty$ (i.e. $L\to\infty$). The excess grand-canonical free energy density 
\begin{equation}
\label{excess}
\omega_s(D,\mu,T)= \lim_{L\to\infty}\left[\frac{\Omega(L,D,T,\mu)}{L^{d-1}}-D\omega_b(T,\mu)\right]\;
\end{equation}
is related to the Casimir force  (per unit area) $F(D,\mu,T)$ via 
\beq 
F(D,\mu,T)=-\frac{\partial \omega_s(D,\mu,T)}{\partial D}\;. \label{Cas_for}
\eeq 
The grand-canonical free energy is evaluated as $\Omega(L,D,T,\mu)=-\beta^{-1}\ln \Xi(L,D,T,\mu)$ and the bulk free energy density $\omega_b(T,\mu)$ is given by $\omega_b(T,\mu)=\lim_{L\to\infty}\frac{1}{L^d}\Omega(L,D=L,T,\mu)$. Using Eq. \eqref{partition}, we may write the excess contribution to the grand potential as 
\begin{equation} 
\label{omegas}
\omega_s(D,\mu,T)=\lim_{L\to\infty}\beta^{-1}D\left[\varphi (\bar{s})-\varphi_b(s_0)\right]\;,
\end{equation}
where 
\begin{equation} 
\label{phi_def}
\varphi (\bar{s})=-\frac{\bar{s}^2}{2 a\beta}+\frac{\mu \bar{s}}{a}-\frac{1}{V}\left[\sum_{\textbf{k}\neq({\bf 0}, k_d)}\sum_{r=1}^{\infty}\frac{1}{r}e^{r(\bar{s}-\beta\tilde{\eps_{\textbf{k}}})}-\sum_{k_d}\log\left(1-e^{\bar{s}-\beta\tilde{\epsilon}_{k_d}}\right)\right]\;,
\end{equation}
$\bar{s}$ denotes the solution to the saddle-point equation $\varphi'(\bar{s})=0$, while $s_0$ corresponds to $\bar{s}$ in the bulk case (i.e. when $D=L$ and $L\to\infty$) and $\varphi_b(s)=\lim_{D\to L}\varphi (s)$. The strategy of the subsequent analysis amounts to solving the saddle point equation $\varphi'(\bar{s})=0$ at finite $D$ and evaluating Eq.~(\ref{omegas}) for the obtained value of $\bar{s}$. This yields the excess grand canonical free energy, from which the Casimir force is obtained via Eq.~(\ref{Cas_for}).
\subsection{Summary of the bulk solution}
Below we summarize the presently relevant features of the system in the thermodynamic limit, where $D=L\to\infty$ (see Refs.~\cite{Jakubczyk_2018, Lebek_2020}). Due to the anisotropic nature of the system, there are two characteristic length scales 
\begin{equation}
\lambda_1 = 2A \, \sqrt{ \pi} \, \sqrt{\beta \, t_0} \qquad \lambda_2=A \, \frac{\pi}{\Gamma(5/4)} \, (\beta \, t)^{1/4}\;
\end{equation} 
related to temperature and playing roles analogous to the thermal de Broglie length in the isotropic case. For convenience, some numerical factors are absorbed in the above definitions. We also introduce the 'thermal volume' parameter: 
\beq
V_T=\lambda_1^{d-m}\lambda_2^m\;.
\eeq
Analysis of the saddle-point equation in the thermodynamic limit\cite{Jakubczyk_2018, Lebek_2020} leads to the following expression for the critical line
\begin{equation}
\mu_c(T)=\frac{a}{V_T}\zeta\Big( \frac{1}{\psi} \Big)\;, 
\label{muc}
\end{equation}
where $\zeta$ denotes the Riemann zeta function and 
\begin{equation}
\frac{1}{\psi}=\frac{d}{2}-\frac{m}{4}\;.
\end{equation}
Note that Eq.~(\ref{muc}) is correct for arbitrary $T$ for the present model, while in a situation involving the full lattice dispersion $\epsilon_{\bf k}$ instead of $\tilde{\epsilon}_{\bf k}$  it would only describe the low-$T$ asymptotics \cite{Jakubczyk_2018}. 
It follows that the critical line obeys the universal power law $\mu_c(T) \sim T^{1/\psi}$. The condensed phase is stable for $\mu>\mu_c(T)$ provided $\frac{1}{\psi}>1$. The condition $\frac{1}{\psi}=1$ determines the lower critical dimension $d_l(m)$ of the system. Note that for the usual isotropic case ($m=0$) one recovers $d_l=2$, for the 'uniaxial' case ($m=1$) $d_l=\frac{5}{2}$, and the largest conceivable value of $d_l$ corresponds to $m=d$, where one obtains $d_l=4$. As shown in Ref.~\cite{Jakubczyk_2018}, the universal bulk properties of the system with given $m$ are closely related to the usual isotropic case in effective dimensionality $d_{\text{eff}}=\frac{2}{\psi}$. This in turn can be connected to the spherical\cite{Berlin_1952} (Berlin-Kac) universality class, or the $N\to\infty$ limit of the $O(N)$ models \cite{Stanley_1968, Moshe_2003}. This correspondence is restricted to bulk properties. 
\section{Saddle-point equation} 
The explicit expression for the saddle-point equation $\varphi'(\bar{s})=0$ can be obtained from Eq.~(\ref{phi_def}), and cast in the following form\cite{Lebek_2020} 
\begin{equation}
\begin{aligned}
\zeta \Big(\epsi\Big) \bigg(-\frac{\bar{s}}{\mu_c \beta}+\varepsilon\bigg)&=g_{\epsi}(e^{\bar{s}})-\zeta \Big(\epsi\Big)+ \frac{1}{\Gamma(5/4)}\mathcal{R}_1^{(1)}+ 
\frac{\Gamma(5/4)^{\frac{4}{\psi}-5}}{\pi^{\frac{4}{\psi} -4}} \Big ( \frac{\lambda_2}{D} \Big)^{\frac{4}{\psi}-4} \sigma^{\frac{4}{\psi}-4} \sum _{n=1}^{\infty} F_{\epsi}(n \sigma) +\\
&-\frac{V_T}{V} \sum_{\vkd} \frac{1 }{1-e^{\beta \tilde{\epsilon}_{\vkd} -\bar{s}}}\;, 
\end{aligned}
\label{SPE2}
\end{equation} 
where $g_n(z)=\sum_{k=1}^\infty \frac{z^k}{k^n}$ denotes the Bose function, $\varepsilon=\frac{\mu -\mu_c}{\mu_c}$, while  
\begin{equation}
\label{sigmadef}
\sigma=\frac{\pi}{\Gamma(5/4)} \frac{D}{\lambda_2} |\bar{s}|^{1/4} \qquad F_{\kappa} (x) = \int_0^{\infty} \di p \, \frac{e^{-p}}{p^\kappa} \phi (x/p^{1/4})
\end{equation} 
and 
\begin{equation}
\label{phi_function}
\phi(k)= \int_{-\infty}^{\infty} \di x \, e^{i k x} \, e^{-x^4}\;
\end{equation}
is the Fourier transform of the quartic Gaussian. Eq.~(\ref{SPE2}) is exact and valid in an arbitrary thermodynamic state, for any $D$ and $L\gg D$. The remainder term $\mathcal{R}_1^{(1)}$ arises from application of the 
Euler-Maclaurin formula\cite{Lebek_2020} and, importantly, can be dropped for $L\to\infty$ (even at $D$ finite). This fact, first demonstrated here, is crucial for the physical conclusions obtained by us in this paper.   % for $D/\lambda_2\gg 1$. 
We provide an analysis of this term in appendix~\ref{appendixA}. Upon neglecting $\mathcal{R}_1^{(1)}$ Eq.~(\ref{SPE2}) becomes equivalent to Eq.~(42) of Ref.~\cite{Lebek_2020}. We continue by introducing the scaling variable 
\begin{equation}
x=\begin{cases}
        \varepsilon \Big(\frac{D}{\lambda_2} \Big)^{\frac{4}{\psi}-4}, &  1<\epsi<2 \\
        \varepsilon \Big(\frac{D}{\lambda_2} \Big)^4, &  \epsi >2\;,
        \end{cases}
\end{equation} 
the sign of which is positive below bulk $T_c$ and negative otherwise. We do not analyze the case $\frac{1}{\psi}=2$ (corresponding to the upper critical dimension of the bulk transition) where $|\bar{s}|$ acquires logarithmic corrections. Note\cite{Jakubczyk_2018} that $x\sim (D/\xi_{\parallel})^\gamma$ with $\gamma=\frac{4}{\psi}-4$ for $1<\epsi<2$ and $\gamma=4$ for $\epsi >2$. The saddle-point equation is finally cast in the following convenient form: %[compare Eq.~(44) and (45) in Ref.~\cite{Lebek_2020}]:
\begin{equation}
\begin{aligned}
\label{eqsadle}
\zeta (\epsi) x =&\bigg(g_{\frac{1}{\psi}}(e^{\bar{s}})-\zeta(\frac{1}{\psi})+\frac{1}{\Gamma(5/4)}\mathcal{R}_1^{(1)}+ \zeta(\frac{1}{\psi})\frac{\bar{s}}{\beta\mu_c}\bigg)\bigg(\frac{D}{\lambda_2}\bigg)^{4(\frac{1}{\psi}-1)}+\\
&\frac{\Gamma(5/4)^{\frac{4}{\psi}-5}}{\pi^{\frac{4}{\psi}-4}}\sigma^{\frac{4}{\psi}-4}\sum_{n=1}^{\infty} F_{\epsi} (n \sigma) + 
\frac{V_T}{V} \left(\frac{D}{\lambda_2}\right)^{\frac{4}{\psi}-4}  \sum_{\vkd} \frac{1 }{e^{\beta \tilde{\epsilon}_{\vkd} -\bar{s}}-1}
\end{aligned}
\end{equation}
for $\epsi <2$ and    
\begin{equation}
\begin{aligned} 
\label{sad1}
\zeta (\epsi) x =  &\bigg(g_{\frac{1}{\psi}}(e^{\bar{s}})-\zeta(\frac{1}{\psi})+\frac{1}{\Gamma(5/4)}\mathcal{R}_1^{(1)}+ \zeta(\frac{1}{\psi})\frac{\bar{s}}{\beta\mu_c}\bigg)\bigg(\frac{D}{\lambda_2}\bigg)^{4}+%\nonumber 
\\
& \frac{\Gamma(5/4)^{\frac{4}{\psi}-5}}{\pi^{\frac{4}{\psi}-4}}\bigg(\frac{\lambda_2}{D}\bigg)^{\frac{4}{\psi}}   \sigma^{\frac{4}{\psi}-4}\sum_{n=1}^{\infty} F_{\epsi} (n \sigma) + \frac{V_T}{V} \left(\frac{D}{\lambda_2}\right)^{4}  \sum_{\vkd} \frac{1 }{e^{\beta \tilde{\epsilon}_{\vkd} -\bar{s}}-1}
\end{aligned}
\end{equation}
for $\epsi >2$.  Note that (up to the remainder terms) Eq.~(\ref{eqsadle}), (\ref{sad1}) correspond to Eq.~(44) and (45) of Ref.~(\cite{Lebek_2020}). We shall now exclusively focus on $x\geq 0$, pertinent to $T\leq T_c$, where the Casimir interaction is expected to be long-ranged. After the above rearrangements, the scaling variable $x$ appears only on the left-hand side of the saddle point equation. At and below criticality $\bar{s}\to 0^-$ and we may expand the right-hand side of the saddle-point equation for $|\bar{s}|\ll 1$.

We now consider the scaling limit, where $\frac{D}{\lambda_2}\gg 1$ and $\varepsilon\ll 1$, while the scaling variable $x$ may take any arbitrary nonnegative value. Recall also that $L\gg D$. In this limit, Eq.~(\ref{eqsadle}) takes the following form: 
\begin{equation} 
\label{spf1}
\zeta (\epsi) x = \bigg(\frac{\Gamma (5/4)}{\pi}\bigg)^{\frac{4}{\psi}-4}\sigma^{\frac{4}{\psi}-4}\left[\Gamma (1-\frac{1}{\psi} )+\frac{1}{\Gamma(5/4)} \sum_{n=1}^{\infty} F_{\epsi} (n \sigma)\right]\
+\frac{V_T}{V|\bar{s}|}\bigg(\frac{D}{\lambda_2}\bigg)^{{\frac{4}{\psi}-4}} +H.O.T.\;,
\end{equation}  
while Eq.~(\ref{sad1}) can be written as
\begin{equation}
\label{spf2}
\zeta (\epsi) x =-\bigg(\frac{\Gamma (5/4)}{\pi}\bigg)^{4}\sigma^{4}\left[\frac{\zeta(\frac{1}{\psi})}{\mu_c\beta}+\zeta(\frac{1}{\psi}-1)\right]+
\frac{V_T}{V|\bar{s}|}\bigg(\frac{D}{\lambda_2}\bigg)^{4} +H.O.T.\;.
\end{equation}
Here '$H.O.T$' stands for terms of higher order in $\bar{s}$ and $\frac{D}{L}$, which do not survive the analyzed limit. 

In both the above cases, the left-hand side (LHS) is positive and does not depend on $|\bar{s}|$. The first term on the right-hand side (RHS) of Eq.~(\ref{spf2}) is manifestly negative. This implies that the second term (involving $V$) must give a finite contribution in the scaling limit to assure the existence of a solution. In consequence, $|\bar{s}|$ is of order $\mathcal{O}(\frac{D^4}{V})$ (assuming the microscopic length scales are of order 1). The situation is similar for Eq.~(\ref{spf1}) provided $\frac{1}{\psi}\geq\frac{7}{4}$ (for $m=1$ this corresponds to $d\geq 4$). We now focus on this case (considering that the opposite situation was analyzed in Ref.~\cite{Lebek_2020}) and inspect Eq.~(\ref{spf1}). Clearly $\Gamma (1-\frac{1}{\psi})<0$, while $F_{\frac{1}{\psi}}$ is bounded from above by its behavior at small arguments (see the appendix~\ref{appendixB}), which is non-positive.  In consequence, for the case described by Eq.~(\ref{spf1}) the last term must give a finite contribution to assure existence of a solution at $x>0$ and  we find   
$|\bar{s}|$ to be necessarily of order $\mathcal{O}(\frac{D^{\frac{4}{\psi}-4}}{V})$. In a compact way we write our result as: 
\beq
\label{sbar}
|\bar{s}|=\mathcal{O}\left(\frac{D^\gamma}{V}\right)\;.
\eeq
Eq.~(\ref{sbar}) constitutes the essential new result of this section. 

The key conclusion of the above analysis is that in the limit $L\gg D\gg l_{mic}$ and for $T\leq T_c$ the behavior of $|\bar{s}|$ is controlled by $L$ rather than $D$. Put in other words, if the limit $L\to \infty$ is performed, keeping $D$ finite, there will be no surviving contribution to $|\bar{s}|$. By virtue of Eq.~(\ref{sigmadef}) the same applies to the quantity $\sigma$. This fact opens wide the way to characterize the excess free energy of the system in the scaling limit, which is done in the next section. 

It is worth emphasizing that Eq.~(\ref{sbar}) holds only at criticality and in the low-$T$ phase (for $T\leq T_c$) and for dimensionality $d$ high enough, namely for $\frac{1}{\psi}\geq\frac{7}{4}$ (corresponding to $d\geq 4$ for $m=1$). %The latter guarantees that condensation occurs even for $D$ finite (i.e. for $~L\to \infty$, but $D=const$). 
If the thermodynamic state is fixed above $T_c$, the magnitude of $|\bar{s}|$ in the limit $L\gg D\gg l_{mic}$ is controlled by the distance from the phase transition (measured by the parameter $\varepsilon$). On the other hand, for $T\leq T_c$, but $\frac{1}{\psi}<\frac{7}{4}$ $|\bar{s}|$ is controlled by $D$ (i.e. $|\bar{s}|$ is finite for $L\to\infty$ with $D$ finite, but vanishes if $D\to \infty$).
In what follows we restrict to the cases, where Eq.~(\ref{sbar}) holds. For an analysis of the opposite situations see Ref.~\cite{Lebek_2020}.

\section{Excess grand  canonical free energy}
The result of Eq.~(\ref{sbar}) greatly simplifies the analysis of Eq.~(\ref{omegas}), leading to the determination of the excess grand canonical free energy $\omega_s$. Considering the limit $L\to\infty$ keeping $D$ finite we may simply put $|\bar{s}|\to 0^+$.  We obtain 
\begin{equation}
\label{omegasf}
\omega_s = - k_B T \frac{\chi^{d-m}{\Delta_m^d}}{D^{2d-m-1}}\;,
\end{equation}
where 
\begin{equation}
\label{Deltaf1}
\Delta_m^d = \frac{\Gamma(5/4)^{\frac{4}{\psi}-1}}{\pi^{\frac{4}{\psi}}}\lim_{\sigma\to 0^+} \sigma^{\frac{4}{\psi}}\sum_{n=1}^{\infty}F_{\frac{1}{\psi}+1}(n\sigma)\;,
\end{equation}
and $\chi=\frac{\lambda_2^2}{\lambda_1}$ is a temperature-independent length scale. Note that the magnitude of the Casimir interaction may be greatly amplified by manipulating this parameter.  Upon putting $\theta_A=\frac{1}{2}$ in Eq.~(\ref{Eq3}) [as pertinent to the present situation - see e.g. Refs.~\cite{Diehl_2002, Essafi_2012}] and identifying $\Gamma\leftrightarrow \chi^{d-m}$ we match Eq.~(\ref{Eq3}) with the derived formula of Eq.~(\ref{omegasf}) and identify $\Delta_m^d$ with the scaling function $\Delta(x)$ of Eq.~(\ref{Eq3}). 

Remarkably, Eq.~(\ref{omegasf}) and Eq.~(\ref{Deltaf1}) hold for \emph{arbitrary} nonnegative value of the scaling variable $x$ and represent the entire expression for $\omega_s$ and not only the asymptotic behavior for $D$ large. The physically relevant new result of the present paper now follows from the analysis of Eq.~(\ref{Deltaf1}) upon changing dimensionality. As we demonstrate below, the sign of  $\Delta_m^d$ is sensitive to the value of $\frac{1}{\psi}$ and therefore may be varied while manipulating $d$ and $m$. When restricting to the 'uniaxial' case $m=1$, we show that  $\Delta_m^d$ is precisely zero for natural even dimensionalities starting from $d=4$. 

We may further simplify Eq.~(\ref{Deltaf1}) extracting the asymptotic behavior of the function $F$ - see the appendix~\ref{appendixB}. Introducing 
\beq 
\label{Gdef}
G(\kappa) = \int_0^\infty \di q q^{4\kappa-5}\phi(q)
\eeq 
we obtain
\beq 
\Delta_m^d = 4 \zeta\bigg(\frac{4}{\psi}\bigg)\frac{\Gamma(5/4)^{\frac{4}{\psi}-1}}{\pi^{\frac{4}{\psi}}}G\bigg(\frac{1}{\psi}+1\bigg)\;.
\label{Deltaf2}
\eeq 
Eq.~(\ref{Deltaf2}) was already contained in Ref.~\cite{Lebek_2020}, but its generality and the rich physical consequences encoded in the properties of the function $G$ were completely neglected in that study, which focused mostly on the physical dimensionality $d=3$. As we demonstrate below, the function $G(\kappa)$ changes sign for 
\beq 
\kappa=\kappa_n=\frac{4n+7}{4} \;\;\textrm{with }\;\; n\in\{0,1,2,\dots\}\;.  
\eeq
The fact that $G(\kappa=\kappa_n)=0$ is proven in the appendix~\ref{appendixB}. Below, in Fig.~1 we provide a plot of $G(\kappa)$ evaluated numerically. 
\begin{figure}[h]
\begin{center}
\label{}
\includegraphics[width=10cm]{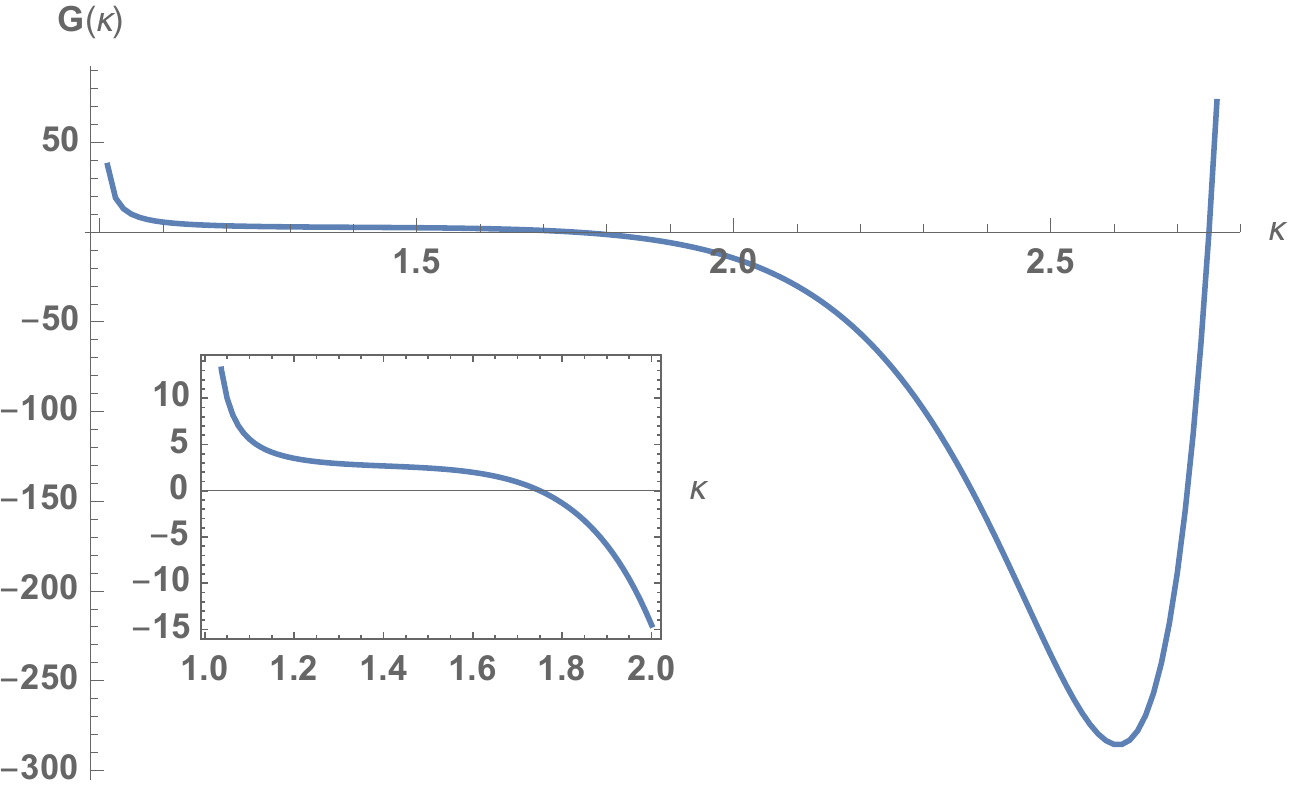}
\caption{The function $G(\kappa)$ evaluated numerically for a range of arguments capturing its first two zeroes (at $\kappa=\kappa_0=\frac{7}{4}$ and  $\kappa=\kappa_1=\frac{11}{4}$). The inset shows the same data in a restricted range of $\kappa$ making the position of $\kappa_0$ clearly visible. For increasing values of $\kappa$, the magnitude of oscillations of $G(\kappa)$ rapidly diverges, so that neither $G(\kappa)$ nor its derivative is bounded from above or below. }
\end{center}
\end{figure} 
Particularly interesting is the case $m=1$, where $\frac{1}{\psi}=\frac{d}{2}-\frac{1}{4}$ and the zeroes of $G(\frac{1}{\psi}+1)$ fall precisely at dimensionality $d=4,6,8,\dots$. In consequence, the Casimir force is repulsive up to dimensionality $d=4$, attractive for $d\in (4,6)$, repulsive again for $d\in (6,8)$ and so on. The associated Casimir amplitude quickly diverges upon increasing $d$. A remarkable observation concerns the case $d\in \{4,6,8,\dots\}$ where the \emph{entire} scaling function $\Delta_{m=1}^d$ is strictly zero, and, as we showed above, there is \emph{no} subleading term surviving the limit $L\to\infty$.  

The equivalence of the presently analyzed model and interacting $N$-component bosons was investigated in Ref.~\cite{Diehl_2017} for $N\to\infty$. In particular the existence of this limit was established therein. One may check, that the form of the dispersion does not influence the validity of the reasoning and the obtained correspondence holds also for anisotropic dispersions. We do not analyze corrections in $1/N$ or the properties of the expansion in the present paper. One can however obtain interesting insights beyond $1/N=0$ imposing only continuity in $1/N$,  
%It is now tempting to speculate about the relevance of the above result beyond $1/N=0$. 
by virtue of which one anticipates a small change of the scaling function when $1/N$ is varied from zero to an arbitrarily small value $1/N=\epsilon$. There is certainly no reason to expect that the dimensionalities marking the boundaries between the attractive and repulsive regimes should still correspond to even natural numbers when $1/N$ is elevated above zero. Nor are   there reasons to anticipate that the scaling function remains constant upon lifting $1/N$. However, considering $1/N$ arbitrarily small, keeping $x$ \emph{fixed} and changing $d$, continuity of the scaling function requires that $\Delta_{m=1}^d(x)$ changes sign at some $d$, which may (and presumably does) depend on $x$.  

 In Fig.~2 we have illustrated the picture we obtained at $1/N=0$ together with the anticipated features at $1/N$ sufficiently small. It is absolutely open, what survives out of this in the physically most interesting cases of $N=3$ and $N=2$, for example whether the dashed lines (boundaries between the repulsive and attractive regimes)  emerging from $1/N=0$ and $d=4,6,8,\dots$ persist up to $1/N=0.3(3)$, or (for example) merge in pairs at some values of $1/N$. Addressing this question is an interesting (presumably challenging) topic for future research.   
\begin{figure}[ht]
\begin{center}
\label{}
\includegraphics[width=9.5cm]{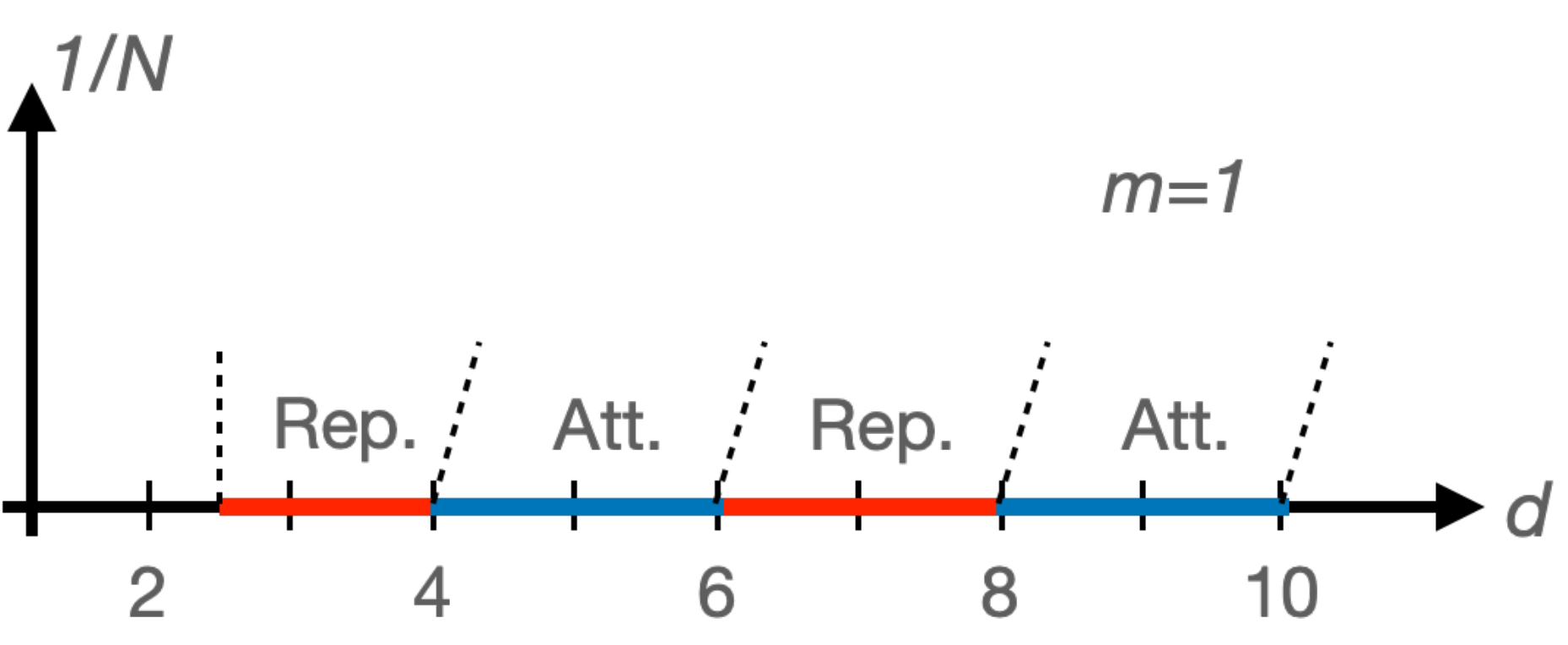}
\caption{Illustration of the results concerning the sign of the Casimir force for $1/N=0$ for the uniaxial case $m=1$ together with the expected  situation at $1/N\ll 1$. At $1/N=0$ the obtained interaction is repulsive for dimensionality $d<4$, attractive for $d\in (4,6)$, repulsive for $d\in (6,8)$ and so on. Occurrence of the boundaries separating the attractive and repulsive regimes (at fixed $x$ - see the main text) is very likely to persist for small $1/N>0$ but their fate upon increasing $1/N$ towards $1/3$ is completely open.     }
\end{center}
\end{figure} 
\section{Discussion and outlook}
In this paper we disclose the surprising properties regarding the sign of the Casimir force in anisotropic systems upon varying dimensionality. Employing the imperfect (mean-field) Bose gas as a representative of the (anisotropic) vectorial $N\to\infty$ class, we demonstrate the periodic alternation of the sign of the Casimir energy upon changing dimensionality. Particularly interesting is the case $m=1$, where the dispersion is quartic in one of the spatial directions and quadratic in the remaining ones. In this situation we demonstrate that the Casimir interaction is repulsive for dimensionality $d\in (\frac{5}{2},4)\cup (6,8)\cup (10,12)\cup\dots$ and attractive for $d\in (4,6)\cup (8,10)\cup \dots$. We show moreover, that for $d\in\{4,6,8\dots\}$ the Casimir interaction entirely vanishes in the scaling limit. Even though the analysis is performed for a system from the $1/N=0$ universality class, from continuity in $1/N$ one may argue that the uncovered unexpected features should also occur for $1/N$ finite but sufficiently small. The phrase 'sufficiently small' is vague here, and this is by no means excluded that the physically most interesting situations of $N\in\{1,2,3\}$ fall into this category. A clarification of this issue requires further studies e.g. from the point of view of the field-theoretic approaches with $1/N$ expansion, or numerical simulations. 

The possibility of modifying the sign of Casimir-type forces (for example by manipulating the boundary fields) was recently considered in a number of contexts\cite{Nowakowski_2008, Abraham_2010, Dohm_2011, Rajabpour_2016, Jakubczyk_2016_2, Sadhukhan_2017, Flachi_2017, Faruk_2018, Voronina_2019, Voronina_2019_2}. The presently analyzed setup predicting their oscillations as function of dimensionality and actual \emph{vanishing} at even values of $d$ appears however entirely new and somewhat surprising.  

We do not expect that our predictions appear at this point obvious for experimental tests considering the magnitude of the considered effects and the high degree of required tuning. They are however certainly open to verification by numerical simulations. Restricting to natural $d$ (where simulations are usually performed) and $m=1$ we then expect a repulsive interaction at $d=3$ and attractive for $d=5$, while the obtained force should entirely vanish in the vicinity of $d=4$. This should be exact provided $1/N$ is sufficiently small. Even though our original motivation stems from cold-atom systems in optical lattices, numerical simulations may, by virtue of universality, be performed using any convenient representative of the universality class. For lattice magnetic systems a paradigmatic choice characterized by $m=1$ might be the Lifshitz point of the so-called ANNNI (anisotropic next-nearest-neighbor Ising) model \cite{Selke_1988} and its counterparts involving a larger number of magnetization components. 

Apart from bosons in optical lattices and thermal phase transitions of the Lifshitz type, our study may be of relevance in the context of quantum phase transitions. This is worth attention, since the scaling properties of a system in the vicinity of many thermal phase transitions are closely related to those of quantum (i.e. occurring at $T=0$ and therefore driven by quantum fluctuations) phase transitions in elevated dimensionality $d_q=d+z$\cite{Sachdev_book}. This equivalence was in particular demonstrated for the isotropic variant of the present model\cite{Jakubczyk_2013_2}. To which extent the quantum-classical correspondence also holds for anisotropic systems and for interfacial properties needs clarifying studies. Assuming such an extension is possible, the peculiar case of $d_q=4$, where the Casimir interaction vanishes should relate to physical dimensionality $d=3$ for systems with a gap in the ordered phase (where $z=1$). On the other hand, interacting bosons are characterized by $z=2$, which makes both $d_q=4$ and $d_q=5$ ($d=2$ and $d=3$ respectively) physically conceivable. The quantum-classical crossover occurring in the vicinity of the quantum critical point, might then be reflected in a crossover between two effective dimensionalities displaying, for example, different sign of the Casimir amplitude.  
An avenue to realize a quantum Lifshitz point was, for example, recently exposed \cite{Zdybel_2020} for imbalanced Fermi mixtures, exhibiting competition between conventional $s$-wave and nonuniform (Fulde-Ferrell-Larkin-Ovchinnikov) \cite{fulde_superconductivity_1964, larkin_nonuniform_1965} superfluid phases.

 %Another context, where our study may be of relevance concerns quantum phase transitions. The present analysis is performed in the vicinity (or below) a thermal phase transition, where $T$ is assumed finite. Technically this means that $D\gg \lambda_{1,2}$, which forbids taking the $T\to 0$ limit. It is however recognized, that (at least bulk) scaling properties of a system in the vicinity of many thermal phase transitions are closely related to those of a quantum (i.e. occurring at $T=0$ and therefore driven by quantum fluctuations) phase transition in elevated dimensionality $d_q=d+z$.\cite{Sachdev_book} This equivalence was in particular demonstrated for the isotropic variant of the present model.\cite{Jakubczyk_2013_2} To which extent the quantum-classical correspondence also holds for anisotropic systems and for interfacial properties needs clarifying studies. Assuming such an extension is possible, the peculiar case of $d_q=4$, where the Casimir interaction vanishes should relate to physical dimensionality $d=3$ for systems with a gap in the ordered phase (where $z=1$). The celebrated example is the quantum Ising model in a transverse field. Its classical counterpart may realize a Lifshitz point upon introducing the next-nearest-neighbor interaction (the so-called ANNNI model).\cite{Selke_1988} On the other hand, interacting bosons are characterized by $z=2$, which makes both $d_q=4$ and $d_q=5$ ($d=2$ and $d=3$ respectively) physically conceivable in anisotropic optical lattice systems.\cite{Greif_2013, Imriska_2014}   

\section*{Acknowledgements}
We are grateful to Marek Napi\'{o}rkowski and Piotr Nowakowski for discussions as well as reading the manuscript and providing helpful suggestions.
\paragraph{Funding information}
PJ acknowledges support from the Polish National Science Center via  2017/26/E/ST3/00211.

\begin{appendix}

\section{The remainder term $\mathcal{R}_1^{(1)}$}
\label{appendixA}
In this appendix we provide a discussion of the remainder term $\mathcal{R}_1^{(1)}$ occurring  in Eq.~(\ref{SPE2}). This arises\cite{Lebek_2020} while approximating the sum $\sum_{r=1}^{\infty} f(r)$ with the integral $\int_0^{\infty}drf(r)$ for 
\beq 
f(r)= e^{r\bar{s}}r^{-\frac{1}{\psi}}\phi (\mathcal{C}_nr^{-\frac{1}{4}})
\eeq
 and $\mathcal{C}_n=\frac{\pi n}{\Gamma(5/4)}\frac{D}{\lambda_2}$ with $n\in\mathbb{N}$. The Euler-Maclaurin formula for our case may be written as
\begin{align} 
\label{LMformula}
\sum_{r=1}^{M}f(r)-\int_0^{M}\di r f(r)&=\frac{f(M)-f(0)}{2}+ \nonumber \\
&\sum_{k=1}^{[p/2]}\frac{b_{2k}}{(2k)!}\left(f^{(2k-1)}(M)-f^{(2k-1)}(0)\right)+\mathcal{R}_p\;,
\end{align}
with $M\to\infty$, $p\in\mathbb{N}$ arbitrary, $[p/2]$ denoting the integer part of $p/2$, $b_{2k}$ being numerical coefficients of no relevance here, and finally
\beq 
\mathcal{R}_p = (-1)^{p+1}\frac{1}{p!}\int_0^M \di r f^{(p)}(r)P_p(r)\;.
\eeq
Here $P_p(r)$ are the periodized Bernoulli functions. The asymptotic forms of the function $\phi(x)$ are given in Ref.~\cite{Boyd_2014}. For the present analysis it is sufficient to know that $\phi(x=0)=const$ and can be expanded in powers of $x$ in the neighborhood of $x=0$, while $|\phi (\mathcal{C}_nr^{-1/4})|\approx r^\gamma e^{-\alpha /r^{1/3}}$ for $r\to 0^+$ with $\gamma$ and $\alpha$ positive. Using these forms for $r\ll1$ and $r\gg 1$ in the definition of $f(r)$ one finds that $f(r)$ vanishes at $r\to 0$ and $r\to\infty$ together with \emph{all} of its derivatives. This is true for any values of $D$. In consequence, $\mathcal{R}_p$ is the only nonzero component on the RHS of Eq.~(\ref{LMformula}). By choosing $p=1$ and recalling that $P_1(r)=\left(r-[r]\right)-\frac{1}{2}$, we obtain that
\beq
\sum_{r=1}^{M}f(r)-\int_0^{M}\di r f(r) =\mathcal{R}_1= \int_0^\infty \di r \left[\frac{d}{dr}\left(\frac{e^{r\bar{s}}}{r^{1/\psi}}\phi\left(\mathcal{C}_nr^{-1/4}\right)\right)\right]P_1(r)\;.
\eeq
We now introduce $x=r|\bar{s}|$, and write $\mathcal{R}_1$ as 
\beq 
\label{rem_fin}
\mathcal{R}_1=|\bar{s}|^{1/\psi}\int_0^\infty \di x \left[\frac{d}{dx}\left(\frac{e^{-x}}{x^{1/\psi}}\phi\left(\frac{n\sigma}{x^{1/4}}\right)\right)\right] \left\{\frac{x}{|\bar{s}|}-\left[\frac{x}{|\bar{s}|}\right]-\frac{1}{2}\right\}\;.
\eeq
The integral is convergent for any $\bar{s}$, and therefore $|\mathcal{R}_1|\leq \mathcal{O}(|\bar{s}|^{1/\psi}$), which is already sufficient to justify dropping the remainder in Eq.~(\ref{SPE2}). The bound we used is very crude, and in fact the integral vanishes for $|\bar{s}|\to 0$ due to the violent oscillations of the term $\left\{\frac{x}{|\bar{s}|}-\left[\frac{x}{|\bar{s}|}\right]-\frac{1}{2}\right\}$.
\section{Properties of the $F$ and $G$ functions}
\label{appendixB}
In this appendix we exhibit the relevant properties of the functions $F$ and $G$. In particular we demonstrate the zeroes of $G$. 

The function $F_\kappa(x)$ is defined in Eq.~(\ref{sigmadef}) and a change of variables brings it to the form 
\begin{equation}
\label{F_app}
F_{\kappa}(x)= \int_0^{\infty} \di p \, \frac{e^{-p}}{p^{\kappa}} \phi \Big(\frac{x}{p^{1/4}}\Big)=\frac{4}{x^{4\kappa-4}} \int_0^{\infty} \di q \, q^{4\kappa-5} \, e^{-\frac{x^4}{q^4}} \,\phi(q)\;.
\end{equation}
We are interested here only in $\kappa> 1$ and sufficiently small $x$. The integrand on the RHS of Eq.~(\ref{F_app}) is exponentially suppressed for $q\ll x$, while for $q\gg x$ the term $e^{-\frac{x^4}{q^4}}$ may be approximated by unity. Since $\phi (q)$ is constant for $q\to 0$, we may replace $e^{-\frac{x^4}{q^4}}\to 1$ for $x\ll1$. We obtain 
\begin{equation}
%F_{\kappa}(x) \approx 
F_{\kappa}(x\approx 0)\approx\frac{4}{x^{4\kappa-4}}G(\kappa)\;+\dots\;,
\end{equation}

%From the above form, $|F_\kappa(x)|$ is manifestly a decreasing function of $x$, therefore bounded from above by its behavior near $x=0$. It follows that 
%\begin{equation}
%|F_{\kappa}(x)| \leq |F_{\kappa}(x\approx 0)|=\frac{4}{x^{4\kappa-4}}|G(\kappa)|\;+\dots\;,
%\end{equation}
where the function $G(\kappa)$ is defined by Eq.~(\ref{Gdef}). We have checked that for $\kappa\in [\frac{7}{4},2)$ $F_\kappa(x)$ is a decreasing function of $x$, bounded from above by its behavior near $x=0$. 

We now demonstrate that $G(\kappa =\kappa_n)=0$ for $\kappa_n=\frac{4n+7}{4}$ and $n\in\{0,1,2,\dots\}$. Plugging $\kappa =\kappa_n$ into the definition of $G(\kappa)$ and using $\phi(-q)=\phi(q)$ we change the order of the integrals occurring in $G(\kappa)$ and obtain: 
\beq
G(\kappa_n) = \frac{1}{2}\int_{-\infty}^{\infty} \di x e^{-x^4}\int_ {-\infty}^{\infty} \di q q^{4n+2}e^{iqx}\;.
\eeq
We now use the representation of the $l$-th derivative of Dirac delta: 
\beq
\delta^{(l)}(x)= \frac{1}{2\pi} \int_ {-\infty}^{\infty} \di q (iq)^{l} e^{iqx}\;,
\eeq
which leads to 
\beq
G(\kappa_n) = (-i)^{4n+2}\pi\int_{-\infty}^{\infty} \di x \delta^{(4n+2)}(x)e^{-x^4}\;.
\eeq
Considering that $\int_{-\infty}^{\infty} \di x\delta^{(l)}(x)f(x)=(-1)^l f^{(l)}(0)$, we find:
\beq 
G(\kappa_n)= \pi (i)^{4n+2} \left(e^{-x^4}\right)^{(4n+2)}|_{x=0} = 0\;.
\eeq
In performing the last step we observed, that only the 4-th, 8-th, 12-th and so on derivatives of the function $f(x)=e^{-x^4}$ are nonvanishing at $x=0$. 

%\section{Relation to the classical field theory in the limit $N\to\infty$}
%\label{appendix c}
%Below we explain the relation between the imperfect Bose gas and the classical field theory for an $N$-component order-parameter field in the limit $N\to\infty$. Our discussion may be viewed as an adaptation of the analysis performed in Ref.~\cite{Diehl_2017} to systems with arbitrary (anisotropic) dispersions, which, however, does not bring in conceptually new elements. 
%\begin{equation} 
%S[]=\int_k \psi^*\left(\hbar i \omega +\epsilon_{{\bf k}}-\mu  \right)\psi\;.
%\end{equation}

\end{appendix}

% TODO:
% Provide your bibliography here. You have two options:

% FIRST OPTION - write your entries here directly, following the example below, including Author(s), Title, Journal Ref. with year in parentheses at the end, followed by the DOI number.
%\begin{thebibliography}{99}
%\bibitem{1931_Bethe_ZP_71} H. A. Bethe, {\it Zur Theorie der Metalle. i. Eigenwerte und Eigenfunktionen der linearen Atomkette}, Zeit. f{\"u}r Phys. {\bf 71}, 205 (1931), \doi{10.1007\%2FBF01341708}.
%\bibitem{arXiv:1108.2700} P. Ginsparg, {\it It was twenty years ago today... }, \url{http://arxiv.org/abs/1108.2700}.
%\end{thebibliography}

% SECOND OPTION:
% Use your bibtex library
%\bibliographystyle{SciPost_bibstyle} % Include this style file here only if you are not using our template
\bibliography{SciPost_Example_BiBTeX_File.bib}

\begin{thebibliography}{10}
\providecommand{\url}[1]{\texttt{#1}}
\providecommand{\urlprefix}{URL }
\expandafter\ifx\csname urlstyle\endcsname\relax
  \providecommand{\doi}[1]{doi:\discretionary{}{}{}#1}\else
  \providecommand{\doi}{doi:\discretionary{}{}{}\begingroup
  \urlstyle{rm}\Url}\fi
\providecommand{\eprint}[2][]{\url{#2}}

\bibitem{Mostepanenko_1988}
V.~M. Mostepanenko and N.~N. Trunov,
\newblock \emph{The {C}asimir effect and its applications},
\newblock Soviet Physics Uspekhi \textbf{31}(11), 965 (1988),
\newblock \doi{10.1070/pu1988v031n11abeh005641}.

\bibitem{Krech_book}
M.~Krech,
\newblock \emph{The Casimir Effect in Critical Systems},
\newblock WORLD SCIENTIFIC,
\newblock \doi{10.1142/2434} (1994).

\bibitem{Kardar_1999}
M.~Kardar and R.~Golestanian,
\newblock \emph{The ``friction'' of vacuum, and other fluctuation-induced
  forces},
\newblock Rev. Mod. Phys. \textbf{71}, 1233 (1999),
\newblock \doi{10.1103/RevModPhys.71.1233}.

\bibitem{Brankov_book}
J.~G. Brankov, D.~M. Danchev and N.~S. Tonchev,
\newblock \emph{Theory of Critical Phenomena in Finite-Size Systems},
\newblock WORLD SCIENTIFIC,
\newblock \doi{10.1142/4146} (2000).

\bibitem{Dantchev_2003}
D.~Dantchev, M.~Krech and S.~Dietrich,
\newblock \emph{Universality of the thermodynamic {C}asimir effect},
\newblock Phys. Rev. E \textbf{67}, 066120 (2003),
\newblock \doi{10.1103/PhysRevE.67.066120}.

\bibitem{Gambassi_2009}
A.~Gambassi,
\newblock \emph{{The Casimir effect: From quantum to critical fluctuations}},
\newblock J. Phys. Conf. Ser. \textbf{161}, 012037 (2009),
\newblock \doi{10.1088/1742-6596/161/1/012037},
\newblock \eprint{0812.0935}.

\bibitem{Klimchitskaya_2009}
G.~L. Klimchitskaya, U.~Mohideen and V.~M. Mostepanenko,
\newblock \emph{The {C}asimir force between real materials: {E}xperiment and
  theory},
\newblock Rev. Mod. Phys. \textbf{81}, 1827 (2009),
\newblock \doi{10.1103/RevModPhys.81.1827}.

\bibitem{Maciolek_2018}
A.~Macio\l{}ek and S.~Dietrich,
\newblock \emph{Collective behavior of colloids due to critical {C}asimir
  interactions},
\newblock Rev. Mod. Phys. \textbf{90}, 045001 (2018),
\newblock \doi{10.1103/RevModPhys.90.045001}.

\bibitem{Machta_2011}
B.~B. Machta, S.~Papanikolaou, J.~P. Sethna and S.~L. Veatch,
\newblock \emph{Minimal model of plasma membrane heterogeneity requires
  coupling cortical actin to criticality},
\newblock Biophys J. \textbf{100}, 1668 (2011),
\newblock \doi{10.1016/j.bpj.2011.02.029}.

\bibitem{Machta_2012}
B.~B. Machta, S.~L. Veatch and J.~P. Sethna,
\newblock \emph{Critical {C}asimir {F}orces in {C}ellular {M}embranes},
\newblock Phys. Rev. Lett. \textbf{109}, 138101 (2012),
\newblock \doi{10.1103/PhysRevLett.109.138101}.

\bibitem{Vasilyev_2009}
O.~Vasilyev, A.~Gambassi, A.~Macio\l{}ek and S.~Dietrich,
\newblock \emph{Universal scaling functions of critical {C}asimir forces
  obtained by {M}onte {C}arlo simulations},
\newblock Phys. Rev. E \textbf{79}, 041142 (2009),
\newblock \doi{10.1103/PhysRevE.79.041142}.

\bibitem{Li_1997}
X.-z. Li, H.-b. Cheng, J.-m. Li and X.-h. Zhai,
\newblock \emph{Attractive or repulsive nature of the {C}asimir force for
  rectangular cavity},
\newblock Phys. Rev. D \textbf{56}, 2155 (1997),
\newblock \doi{10.1103/PhysRevD.56.2155}.

\bibitem{Kenneth_2006}
O.~Kenneth and I.~Klich,
\newblock \emph{Opposites {A}ttract: {A} {T}heorem about the {C}asimir
  {F}orce},
\newblock Phys. Rev. Lett. \textbf{97}, 160401 (2006),
\newblock \doi{10.1103/PhysRevLett.97.160401}.

\bibitem{Soyka_2008}
F.~Soyka, O.~Zvyagolskaya, C.~Hertlein, L.~Helden and C.~Bechinger,
\newblock \emph{Critical {C}asimir {F}orces in {C}olloidal {S}uspensions on
  {C}hemically {P}atterned {S}urfaces},
\newblock Phys. Rev. Lett. \textbf{101}, 208301 (2008),
\newblock \doi{10.1103/PhysRevLett.101.208301}.

\bibitem{Nellen_2009}
U.~Nellen, L.~Helden and C.~Bechinger,
\newblock \emph{Tunability of critical {C}asimir interactions by boundary
  conditions},
\newblock {EPL} (Europhysics Letters) \textbf{88}(2), 26001 (2009),
\newblock \doi{10.1209/0295-5075/88/26001}.

\bibitem{Greif_2013}
D.~Greif, T.~Uehlinger, G.~Jotzu, L.~Tarruell and T.~Esslinger,
\newblock \emph{Short-{R}ange {Q}uantum {M}agnetism of {U}ltracold {F}ermions
  in an {O}ptical {L}attice},
\newblock Science \textbf{340}(6138), 1307 (2013),
\newblock \doi{10.1126/science.1236362}.

\bibitem{Imriska_2014}
J.~Imri\ifmmode~\check{s}\else \v{s}\fi{}ka, M.~Iazzi, L.~Wang, E.~Gull,
  D.~Greif, T.~Uehlinger, G.~Jotzu, L.~Tarruell, T.~Esslinger and M.~Troyer,
\newblock \emph{Thermodynamics and {M}agnetic {P}roperties of the {A}nisotropic
  {3D} {H}ubbard {M}odel},
\newblock Phys. Rev. Lett. \textbf{112}, 115301 (2014),
\newblock \doi{10.1103/PhysRevLett.112.115301}.

\bibitem{Jakubczyk_2018}
P.~Jakubczyk and J.~Wojtkiewicz,
\newblock \emph{Phase diagram and correlation functions of the anisotropic
  imperfect {B}ose gas in {$d$} dimensions},
\newblock Journal of Statistical Mechanics: Theory and Experiment
  \textbf{2018}(5), 053105 (2018),
\newblock \doi{10.1088/1742-5468/aabc7c}.

\bibitem{Lebek_2020}
M.~\L{}ebek and P.~Jakubczyk,
\newblock \emph{Dimensional crossovers and {C}asimir forces for the {B}ose gas
  in anisotropic optical lattices},
\newblock Phys. Rev. A \textbf{102}, 013324 (2020),
\newblock \doi{10.1103/PhysRevA.102.013324}.

\bibitem{Dantchev_2006}
D.~Dantchev, H.~W. Diehl and D.~Gr\"uneberg,
\newblock \emph{Excess free energy and {C}asimir forces in systems with
  long-range interactions of van der {W}aals type: {G}eneral considerations and
  exact spherical-model results},
\newblock Phys. Rev. E \textbf{73}, 016131 (2006),
\newblock \doi{10.1103/PhysRevE.73.016131}.

\bibitem{Nowakowski_2008}
P.~Nowakowski and M.~Napi\'orkowski,
\newblock \emph{Scaling of solvation force in two-dimensional {I}sing strips},
\newblock Phys. Rev. E \textbf{78}, 060602 (2008),
\newblock \doi{10.1103/PhysRevE.78.060602}.

\bibitem{Nowakowski_2009}
P.~Nowakowski and M.~Napi{\'{o}}rkowski,
\newblock \emph{Properties of the solvation force of a two-dimensional {I}sing
  strip in scaling regimes},
\newblock Journal of Physics A: Mathematical and Theoretical \textbf{42}(47),
  475005 (2009),
\newblock \doi{10.1088/1751-8113/42/47/475005}.

\bibitem{Dantchev_2011}
J.~Bergknoff, D.~Dantchev and J.~Rudnick,
\newblock \emph{Casimir force in the rotor model with twisted boundary
  conditions},
\newblock Phys. Rev. E \textbf{84}, 041134 (2011),
\newblock \doi{10.1103/PhysRevE.84.041134}.

\bibitem{Diehl_2012}
H.~W. Diehl, D.~Grüneberg, M.~Hasenbusch, A.~Hucht, S.~B. Rutkevich and F.~M.
  Schmidt,
\newblock \emph{Exact thermodynamic casimir forces for an interacting
  three-dimensional model system in film geometry with free surfaces},
\newblock {EPL} (Europhysics Letters) \textbf{100}(1), 10004 (2012),
\newblock \doi{10.1209/0295-5075/100/10004}.

\bibitem{Hasenbusch_2012}
M.~Hasenbusch,
\newblock \emph{Thermodynamic {C}asimir effect: {U}niversality and corrections
  to scaling},
\newblock Phys. Rev. B \textbf{85}, 174421 (2012),
\newblock \doi{10.1103/PhysRevB.85.174421}.

\bibitem{Hasenbusch_2013}
M.~Hasenbusch,
\newblock \emph{Thermodynamic {C}asimir forces between a sphere and a plate:
  {M}onte {C}arlo simulation of a spin model},
\newblock Phys. Rev. E \textbf{87}, 022130 (2013),
\newblock \doi{10.1103/PhysRevE.87.022130}.

\bibitem{Vasilyev_2013}
O.~A. Vasilyev and S.~Dietrich,
\newblock \emph{Critical {C}asimir forces for films with bulk ordering fields},
\newblock {EPL} (Europhysics Letters) \textbf{104}(6), 60002 (2013),
\newblock \doi{10.1209/0295-5075/104/60002}.

\bibitem{Diehl_2014}
H.~W. Diehl, D.~Gr\"uneberg, M.~Hasenbusch, A.~Hucht, S.~B. Rutkevich and F.~M.
  Schmidt,
\newblock \emph{Large-$n$ approach to thermodynamic {C}asimir effects in slabs
  with free surfaces},
\newblock Phys. Rev. E \textbf{89}, 062123 (2014),
\newblock \doi{10.1103/PhysRevE.89.062123}.

\bibitem{Dantchev_2014}
D.~Dantchev, J.~Bergknoff and J.~Rudnick,
\newblock \emph{Casimir force in the
  $\mathrm{O}(\mathit{n}\ensuremath{\rightarrow}\ensuremath{\infty})$ model
  with free boundary conditions},
\newblock Phys. Rev. E \textbf{89}, 042116 (2014),
\newblock \doi{10.1103/PhysRevE.89.042116}.

\bibitem{Hasenbusch_2015}
M.~Hasenbusch,
\newblock \emph{Thermodynamic {C}asimir effect in films: The exchange cluster
  algorithm},
\newblock Phys. Rev. E \textbf{91}, 022110 (2015),
\newblock \doi{10.1103/PhysRevE.91.022110}.

\bibitem{Diehl_2002}
H.~W. Diehl,
\newblock \emph{Critical behavior at m-axial {L}ifshitz points},
\newblock Acta Physica Slovaca \textbf{52}, 271 (2002).

\bibitem{Burgsmuller_2010}
M.~Burgsmüller, H.~W. Diehl and M.~A. Shpot,
\newblock \emph{Fluctuation-induced forces in strongly anisotropic critical
  systems},
\newblock Journal of Statistical Mechanics: Theory and Experiment
  \textbf{2010}(11), P11020 (2010),
\newblock \doi{10.1088/1742-5468/2010/11/p11020}.

\bibitem{Hornreich_1975}
R.~M. Hornreich, M.~Luban and S.~Shtrikman,
\newblock \emph{Critical behavior at the onset of
  $\stackrel{\ensuremath{\rightarrow}}{\mathrm{k}}$-space instability on the
  $\ensuremath{\lambda}$ line},
\newblock Phys. Rev. Lett. \textbf{35}, 1678 (1975),
\newblock \doi{10.1103/PhysRevLett.35.1678}.

\bibitem{Chaikin_book}
P.~M. Chaikin and T.~C. Lubensky,
\newblock \emph{Principles of condensed matter physics},
\newblock Cambridge University Press (1995).

\bibitem{Diehl_2017}
H.~W. Diehl and S.~B. Rutkevich,
\newblock \emph{Fluctuation-induced forces in confined ideal and imperfect
  {B}ose gases},
\newblock Phys. Rev. E \textbf{95}, 062112 (2017),
\newblock \doi{10.1103/PhysRevE.95.062112}.

\bibitem{Dantchev_1996}
D.~Danchev,
\newblock \emph{Finite-size scaling {C}asimir force function: {E}xact
  spherical-model results},
\newblock Phys. Rev. E \textbf{53}, 2104 (1996),
\newblock \doi{10.1103/PhysRevE.53.2104}.

\bibitem{Dantchev_2004}
D.~Dantchev and M.~Krech,
\newblock \emph{Critical {C}asimir force and its fluctuations in lattice spin
  models: {E}xact and {M}onte {C}arlo results},
\newblock Phys. Rev. E \textbf{69}(4), 046119 (2004),
\newblock \doi{10.1103/PhysRevE.69.046119}.

\bibitem{Kac_1963}
M.~Kac, G.~E. Uhlenbeck and P.~C. Hemmer,
\newblock \emph{On the van der {W}aals {T}heory of the {V}apor‐{L}iquid
  {E}quilibrium. {I}. {D}iscussion of a {O}ne‐{D}imensional {M}odel},
\newblock Journal of Mathematical Physics \textbf{4}(2), 216 (1963),
\newblock \doi{10.1063/1.1703946},
\newblock \eprint{https://doi.org/10.1063/1.1703946}.

\bibitem{Davies_1972}
E.~B. Davies,
\newblock \emph{The thermodynamic limit for an imperfect {B}oson gas},
\newblock Comm. Math. Phys. \textbf{28}(1), 69 (1972).

\bibitem{Buffet_1983}
E.~Buffet and J.~V. Pulè,
\newblock \emph{Fluctuation properties of the imperfect {B}ose gas},
\newblock Journal of Mathematical Physics \textbf{24}(6), 1608 (1983),
\newblock \doi{10.1063/1.525855},
\newblock \eprint{https://doi.org/10.1063/1.525855}.

\bibitem{Zagrebnov_2001}
V.~A. Zagrebnov and J.-B. Bru,
\newblock \emph{The {B}ogoliubov model of weakly imperfect {B}ose gas},
\newblock Physics Reports \textbf{350}(5), 291  (2001),
\newblock \doi{https://doi.org/10.1016/S0370-1573(00)00132-0}.

\bibitem{Napiorkowski_2011}
M.~Napi\'orkowski and J.~Piasecki,
\newblock \emph{Casimir force induced by an imperfect {B}ose gas},
\newblock Phys. Rev. E \textbf{84}, 061105 (2011),
\newblock \doi{10.1103/PhysRevE.84.061105}.

\bibitem{Napiorkowski_2013}
M.~Napi{\'{o}}rkowski, P.~Jakubczyk and K.~Nowak,
\newblock \emph{The imperfect {B}ose gas in {$d$} dimensions: critical behavior
  and {C}asimir forces},
\newblock Journal of Statistical Mechanics: Theory and Experiment
  \textbf{2013}(06), P06015 (2013),
\newblock \doi{10.1088/1742-5468/2013/06/p06015}.

\bibitem{Napiorkowski_2017}
M.~Napi\'orkowski and J.~Piasecki,
\newblock \emph{Thermodynamic equivalence of two-dimensional imperfect
  attractive {F}ermi and repulsive {B}ose gases},
\newblock Phys. Rev. A \textbf{95}, 063627 (2017),
\newblock \doi{10.1103/PhysRevA.95.063627}.

\bibitem{Mysliwy_2019}
K.~My{\'{s}}liwy and M.~Napi{\'{o}}rkowski,
\newblock \emph{Thermodynamics of inhomogeneous imperfect quantum gases in
  harmonic traps},
\newblock Journal of Statistical Mechanics: Theory and Experiment
  \textbf{2019}(6), 063101 (2019),
\newblock \doi{10.1088/1742-5468/ab190d}.

\bibitem{Dantchev_2020}
D.~M. Dantchev,
\newblock \emph{Exact results for the {C}asimir force of a three-dimensional
  model of relativistic {B}ose gas in a film geometry},
\newblock Journal of Statistical Mechanics: Theory and Experiment
  \textbf{2020}(6), 063103 (2020),
\newblock \doi{10.1088/1742-5468/ab900a}.

\bibitem{Ziff_1977}
R.~M. Ziff, G.~E. Uhlenbeck and M.~Kac,
\newblock \emph{The ideal {B}ose-{E}instein gas, revisited},
\newblock Physics Reports \textbf{32}(4), 169  (1977),
\newblock \doi{https://doi.org/10.1016/0370-1573(77)90052-7}.

\bibitem{Berlin_1952}
T.~H. Berlin and M.~Kac,
\newblock \emph{The {S}pherical {M}odel of a {F}erromagnet},
\newblock Phys. Rev. \textbf{86}, 821 (1952),
\newblock \doi{10.1103/PhysRev.86.821}.

\bibitem{Stanley_1968}
H.~E. Stanley,
\newblock \emph{Spherical {M}odel as the {L}imit of {I}nfinite {S}pin
  {D}imensionality},
\newblock Phys. Rev. \textbf{176}, 718 (1968),
\newblock \doi{10.1103/PhysRev.176.718}.

\bibitem{Moshe_2003}
M.~Moshe and J.~Zinn-Justin,
\newblock \emph{Quantum field theory in the large {N} limit: a review},
\newblock Physics Reports \textbf{385}(3), 69  (2003),
\newblock \doi{https://doi.org/10.1016/S0370-1573(03)00263-1}.

\bibitem{Essafi_2012}
{Essafi, K.}, {Kownacki, J.-P.} and {Mouhanna, D.},
\newblock \emph{Nonperturbative renormalization group approach to {L}ifshitz
  critical behaviour},
\newblock EPL \textbf{98}(5), 51002 (2012),
\newblock \doi{10.1209/0295-5075/98/51002}.

\bibitem{Abraham_2010}
D.~B. Abraham and A.~Macio\l{}ek,
\newblock \emph{Casimir {I}nteractions in {I}sing {S}trips with {B}oundary
  {F}ields: {E}xact {R}esults},
\newblock Phys. Rev. Lett. \textbf{105}, 055701 (2010),
\newblock \doi{10.1103/PhysRevLett.105.055701}.

\bibitem{Dohm_2011}
V.~Dohm,
\newblock \emph{Critical free energy and {C}asimir forces in rectangular
  geometries},
\newblock Phys. Rev. E \textbf{84}, 021108 (2011),
\newblock \doi{10.1103/PhysRevE.84.021108}.

\bibitem{Rajabpour_2016}
M.~A. Rajabpour,
\newblock \emph{Classification of the sign of the critical {C}asimir force in
  two-dimensional systems at asymptotically large separations},
\newblock Phys. Rev. D \textbf{94}, 105029 (2016),
\newblock \doi{10.1103/PhysRevD.94.105029}.

\bibitem{Jakubczyk_2016_2}
P.~Jakubczyk, M.~Napi{\'{o}}rkowski and T.~Sek,
\newblock \emph{Repulsive {C}asimir forces at quantum criticality},
\newblock {EPL} (Europhysics Letters) \textbf{113}(3), 30006 (2016),
\newblock \doi{10.1209/0295-5075/113/30006}.

\bibitem{Sadhukhan_2017}
M.~Sadhukhan and A.~Tkatchenko,
\newblock \emph{Long-{R}ange {R}epulsion {B}etween {S}patially {C}onfined van
  der {W}aals {D}imers},
\newblock Phys. Rev. Lett. \textbf{118}, 210402 (2017),
\newblock \doi{10.1103/PhysRevLett.118.210402}.

\bibitem{Flachi_2017}
A.~Flachi, M.~Nitta, S.~Takada and R.~Yoshii,
\newblock \emph{Sign {F}lip in the {C}asimir {F}orce for {I}nteracting
  {F}ermion {S}ystems},
\newblock Phys. Rev. Lett. \textbf{119}, 031601 (2017),
\newblock \doi{10.1103/PhysRevLett.119.031601}.

\bibitem{Faruk_2018}
M.~M. Faruk and S.~Biswas,
\newblock \emph{Repulsive {C}asimir force in {B}ose{\textendash}{E}instein
  {C}ondensate},
\newblock Journal of Statistical Mechanics: Theory and Experiment
  \textbf{2018}(4), 043401 (2018),
\newblock \doi{10.1088/1742-5468/aab01b}.

\bibitem{Voronina_2019}
Y.~Voronina, I.~Komissarov and K.~Sveshnikov,
\newblock \emph{Casimir interactions between two short-range coulomb sources},
\newblock Annals of Physics \textbf{404}, 132  (2019),
\newblock \doi{https://doi.org/10.1016/j.aop.2019.02.014}.

\bibitem{Voronina_2019_2}
Y.~Voronina, I.~Komissarov and K.~Sveshnikov,
\newblock \emph{Casimir force variability in one-dimensional {QED} systems},
\newblock Phys. Rev. A \textbf{99}, 062504 (2019),
\newblock \doi{10.1103/PhysRevA.99.062504}.

\bibitem{Selke_1988}
W.~Selke,
\newblock \emph{The {ANNNI} model — {T}heoretical analysis and experimental
  application},
\newblock Physics Reports \textbf{170}(4), 213  (1988),
\newblock \doi{https://doi.org/10.1016/0370-1573(88)90140-8}.

\bibitem{Sachdev_book}
S.~Sachdev,
\newblock \emph{Quantum Phase Transitions},
\newblock Cambridge University Press, 2 edn.,
\newblock \doi{10.1017/CBO9780511973765} (2011).

\bibitem{Jakubczyk_2013_2}
P.~Jakubczyk and M.~Napi{\'{o}}rkowski,
\newblock \emph{Quantum criticality of the imperfect {B}ose gas in
  {$d$}-dimensions},
\newblock Journal of Statistical Mechanics: Theory and Experiment
  \textbf{2013}(10), P10019 (2013),
\newblock \doi{10.1088/1742-5468/2013/10/p10019}.

\bibitem{Zdybel_2020}
P.~Zdybel and P.~Jakubczyk,
\newblock \emph{Quantum lifshitz points and fluctuation-induced first-order
  phase transitions in imbalanced fermi mixtures},
\newblock Phys. Rev. Research \textbf{2}, 033486 (2020),
\newblock \doi{10.1103/PhysRevResearch.2.033486}.

\bibitem{fulde_superconductivity_1964}
P.~Fulde and R.~A. Ferrell,
\newblock \emph{Superconductivity in a {Strong} {Spin}-{Exchange} {Field}},
\newblock Phys. Rev. \textbf{135}(3A), A550 (1964),
\newblock \doi{10.1103/PhysRev.135.A550}.

\bibitem{larkin_nonuniform_1965}
A.~I. Larkin and Y.~N. Ovchinnikov,
\newblock \emph{Nonuniform state of superconductors},
\newblock Sov. Phys. JETP \textbf{20}, 762 (1965).

\bibitem{Boyd_2014}
J.~P. Boyd,
\newblock \emph{The {F}ourier {T}ransform of the quartic {G}aussian
  {$\exp(-Ax^4)$}: {H}ypergeometric functions, power series, steepest descent
  asymptotics and hyperasymptotics and extensions to {$\exp(-Ax^{2n})$}},
\newblock Applied Mathematics and Computation \textbf{241}, 75  (2014),
\newblock \doi{https://doi.org/10.1016/j.amc.2014.05.001}.

\end{thebibliography}

\nolinenumbers

\end{document}